# Technologies and Workflow of Creative Coding Projects: Examples from the Google DevArt Competition

**Youyang Hou**

Google

1600 Amphitheatre Parkway, CA 94043, USA

youyangh@google.com

## Abstract

Recently, many artists and creative technologists created computer programming as a goal to create something expressive instead of something functional. In this paper, I analyzed 18 creative coding projects from the Google DevArt competition and summarized the critical technologies, types of content, and key workflows of these creative coding projects. The paper also discusses the potential research opportunities for creative coding and art technologies.

## CSS Concepts
• **Human-centered computing~Human computer interaction (HCI)**; Graphical user interface; User interface programming; User interface toolkits

## Introduction
Creative coding is defined as a type of computer programming in which the goal is to create something expressive instead of something functional (https://en.wikipedia.org/wiki/Creative_coding). Creative coding has been considered as a way of teaching programming languages [1, 2, 3]. Creative coding opens up opportunities in new areas of art and design, yet it could hard to understand and execute. To better understand creative coder's needs for technologies and their workflow, I did a content analysis of 18 projects from the Google DevArt contest. Understanding the content, process, and creative coder's needs will help researchers create better creative coding tooling and reduce the barriers for artists to use technologies and programming languages in the art creation process.

## Google DevArt
Google initiated the DevArt project as "a celebration of art made with code by artists that push the possibilities of creativity - where technology is their canvas and code is their raw material". (https://devart.withgoogle.com/). As Google Creative Lab's Emma Turpin said, "What we're trying to show is that art isn't just the output, but the entire development process". Google DevArt contest can help understand creative coders' workflow, technologies, and the relationship between art the technology. Park and Han [4] compared the apps uploaded to Android Marketplace and in Google DevArt contest. They found participants used multiple open platform development tools and languages to create innovative artwork in DevArt projects. However, the content and process of creating DevArt projects have not been well investigated in previous research.

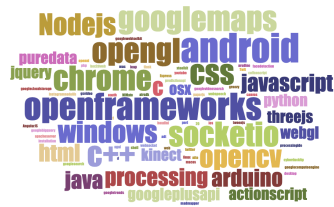

**Figure 1** The word cloud of technologies used in Google DevArt shortlist and finalist projects.

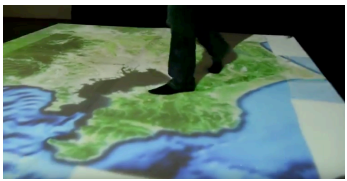

**Figure 2** The Giant Map project

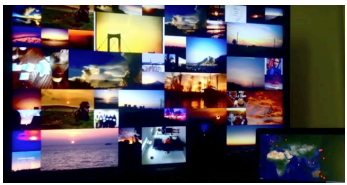

**Figure 3** The Eternal Sunset project

In this paper, I will conduct a content analysis of the top 18 projects from the Google DevArt contest final list and short list winners. All the data I analyzed are publicly available on the Google DevArt website (https://devart.withgoogle.com/).

## Creative Coding Projects and Technologies

All the projects have used multiple technologies (M=7.5, SD=6.36). The most popular technologies included Android (7), Openframework (6), OpenGL (6), Google Maps API (5), CSS (5), C++ (5), and Chrome (5) (Figure 1). Popular creative coding technologies included Openframework, OpenGL, OpenCV, Figure listed a word cloud of technologies being used in the DevArt projects.

There are different categories of art projects. Below I will discuss the characteristics of each type of DevArt projects and the potential for art technology research in the CHI community.

- **Maps**
3 projects used Maps through technologies like Google Maps API, OpenFrameworks, OpenGL, and socketio. For example, the Giant Map project (Figure 2) is an ultra-large interactive Google Map, where children can physically play. KUAFU / 夸父 is an interactive journey through Google Maps, using its topographic data to render in realtime the landscape and to use the Google Maps to navigate. Color of World project displays a color palette of places on Google Maps that is on the opposite side of the earth from the visitor. Google Maps is a resource for real-time location information, color patterns, and potential interactions with exhibition visitors.

- **Social media and crowdsourcing**
Many projects leveraged data from social media or crowdsourced participants. Two projects "Infinite Sunset" and "Eternal Sunset" (Figure 3) represented pictures of sunset from Google search and Instagram. Other projects leveraged text data from social media APIs. Dreamsprawler leverages Google+ API to collect people's dreams and map them into a brainimage model. Freedom of Speech Kit creates a platform that provides access to those who for any reason cannot be on the street to share their opinions and ideas by sending a message through Twitter and an iterative kit.

Current social media technologies provided a variety of image and video sharing APIs such as an attractive prospect for visual inspiration. Social media API also provided rich text content for inspiration of ideas and stories of DevArt projects.

- **Generative art**
Many projects took the format of generative art and transform a traditional art piece into an autogenerative format. Ellsworth Kelly Animated created an animated version of Ellsworth Kelly's paintings. Endemic created a simple biological simulation to represent mutating bacteria. Many also receive visitor's feedback and try to be interactive. For example, the Playful Geometries project (Figure 4) created interactive geometrics patterns which patterns and music changes according to user's input.

- **Multi devices experience**

Many DevArt projects took the format of installation and interaction with visitors. These projects often require cross-platforms development on mobile, web, and hardware. The installation and interactive pieces commonly used Kinect, OpenCV, and RasperryPi. However, many creators failed to create multi-devices experience due to a lack of time and technological experiences.

For instance, the creator of the DreamSprawler project noted in their journal: *"We thought that if we started with the web technology, we could have done a more comprehensive prototype using WebGL, X toolkit, Socket.IO, and others. This is because web applications are mostly faster to develop… We want to create a web interface as well but we realized that coding two platforms in a month is a bit too ambitious. "*

The creator of the Infinite Sunset project, has to give up creating Android app version due to technical difficulties *"I'm also abandoning creating an Android native app, as the Processing IDE is difficult to troubleshoot in Android mode…. things change often in open source world. I'm going to just focus on completing the piece for stand-alone desktop and web."*

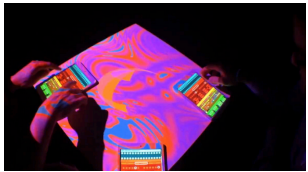

**Figure 4** The Playful Geometries project

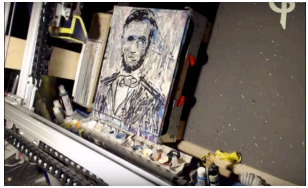

**Figure 5** The Crowd Painter project

Some projects also used physical and tangible materials. The Crowd Painter project (Figure 5) built a painting robot and participants could use a browser or touchscreen to collaborate paint on a canvas.

For creative code projects with multi-devices and multi-screens experience, cross-platform frameworks such as Xamarin, Flutter, and React Native, and Adobe PhoneGap could reduce the development time and enable smooth experience across different devices. How creative coders could leverage these cross-platform technologies is an interesting topic to research in future.

## Workflow

I also analyzed the project journal posted by the creators and found some typical steps when they worked on the creative coding projects: inspiration, sketching, technology researching, data collection, creating algorithms, art effects, prototyping, project testing, installation, and competition. Below I'll discuss some tasks and challenges described by the creators, which could inform future research to further investigate these workflows in more details.

- **Inspirations**

Many projects got inspired by traditional artwork and transformed it into creative coding projects. For example, The Ellsworth Kelly Animated project created an animated version of Ellsworth Kelly's painting. The Endemic project was inspired by a ceramic tile in the Ashmolean Museum. More accessible database of artwork and design patterns, such as Google Arts & Culture and CodePen, are valuable for creative coding projects.

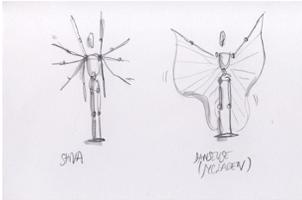

**Figure 6** The paper sketches from the Les métamorphoses de Mr. Kalia project.

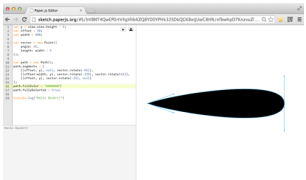

**Figure 7** The paper.js prototype from the Les métamorphoses de Mr. Kalia project.

- **Sketch and Prototype**

Creative coders created various sketches to illustrate their early ideas. Paper or hand-drawn sketches for characteristics, scenes, and installations are very commonly used in these projects (Figure 6). Many projects also used Processing to create digital sketches and prototypes. However, some creators intentionally choose physical sketching and prototyping tools. For example, the creator of The Giant Map said: "*I don't want to use XCode or Processing as an initial step tool. … coding tools and libraries restrict ideas. They define an expression. First I like using various physical tools like paper, photo, model, blocks, etc.*"

Others mentioned the importance of prototyping tools with visual previews (Figure 7). The creator of Les métamorphoses de Mr. Kalia said: "*I've found a great ally: The paper.js sketchpad. It's "just" an online editor providing an interactive quick preview of the sketch. This workflow has proven to be very efficient and iterative since it's easy to share the sketch and get feedback.*"

Future research on tools and technology for art and creative coding could investigate better sketch and prototype tools to enable both creativity and an easy preview of codes and ideas.

- **Art effects**

In the project journals, creators also described their opinions on creating art effects using existing technologies and programming tools. The journals included rich information about how technologies may prevent or facilitate the art creation process. The existing programming tools and technologies were challenging to create certain art effects. For example, the creator of Kuafu noted the look of the experience mix different styles and uses a lot of textures, which is a challenge to recreate in code. Others reflected on the relationship between art and technology. The creator of Palimpsest said: "*The relationship established between art and technology is directly experimental and reflexive … Art asks questions and reflects. Technology emerges as a result of multiple processes of research and scientific elaboration, that in this case, is also proposed as a mediator of the artistic processes, giving to its praxis the possibility of amplifying its experimentation field.*"

- **The Competition**

Google DevArt competition is designed to encourage artists to leverage technologies for creative artwork. Future research should look at how to improve the quality of such competitions. For example, 1) How to reduce the technical barriers of artists in leveraging new technologies and programming languages; 2) How to inspire creative coders and artists by sharing code on Github and document their process.

In addition to the competition, hackathons are also constantly used as way to generate quick creative work within a short period time. For example, University of Colorado organized a Hacking the Gender Gap With Creative Coding (http://atlas.colorado.edu/hacking-the-gender-gap/) Although there has been a lot of research on hackathons for scientific communities [5, 6] and nonprofit organizations [7], future research could investigate the practices of creative coding hackathons and how it leads to creative artworks and better collaborations between artists, designers, and developers.